\documentstyle[12pt]{article} 
\input psfig.sty
\def\Re{{\cal R \mskip-4mu \lower.1ex \hbox{\it e}\,}}
\def\Im{{\cal I \mskip-5mu \lower.1ex \hbox{\it m}\,}}
\def\nn{\noindent}
\def\ie{{\it i.e.}}
\def\eg{{\it e.g.}}

\def\etal{{\it et al.}}

\def\sub#1{_{\lower.25ex\hbox{$\scriptstyle#1$}}}
\def\sul#1{_{\kern-.1em#1}}
\def\sll#1{_{\kern-.2em#1}}  
\def\sbl#1{_{\kern-.1em\lower.25ex\hbox{$\scriptstyle#1$}}}
\def\ssb#1{_{\lower.25ex\hbox{$\scriptscriptstyle#1$}}}
\def\sbb#1{_{\lower.4ex\hbox{$\scriptstyle#1$}}}

\def\to{\rightarrow}
\def\mh{\ifmmode m\sbl H \else $m\sbl H$\fi}
\def\mch{\ifmmode m_{H^\pm} \else $m_{H^\pm}$\fi}
\def\mt{\ifmmode m_t\else $m_t$\fi}
\def\mc{\ifmmode m_c\else $m_c$\fi}
\def\mz{\ifmmode M_Z\else $M_Z$\fi}
\def\mw{\ifmmode M_W\else $M_W$\fi}
\def\mws{\ifmmode M_W^2 \else $M_W^2$\fi}
\def\mhs{\ifmmode m_H^2 \else $m_H^2$\fi}   
\def\mzs{\ifmmode M_Z^2 \else $M_Z^2$\fi}
\def\mts{\ifmmode m_t^2 \else $m_t^2$\fi}
\def\mcs{\ifmmode m_c^2 \else $m_c^2$\fi}
\def\mchs{\ifmmode m_{H^\pm}^2 \else $m_{H^\pm}^2$\fi}
\def\ztwo{\ifmmode Z_2\else $Z_2$\fi}
\def\zone{\ifmmode Z_1\else $Z_1$\fi}
\def\mtwo{\ifmmode M_2\else $M_2$\fi}
\def\mone{\ifmmode M_1\else $M_1$\fi}
\def\tb{\ifmmode \tan\beta \else $\tan\beta$\fi}
\def\xw{\ifmmode x\sub w\else $x\sub w$\fi}
\def\ch{\ifmmode H^\pm \else $H^\pm$\fi}
\def\lum{\ifmmode {\cal L}\else ${\cal L}$\fi}
\def\inpb{\ifmmode {\rm pb}^{-1}\else ${\rm pb}^{-1}$\fi}
\def\infb{\ifmmode {\rm fb}^{-1}\else ${\rm fb}^{-1}$\fi}
\def\epem{\ifmmode e^+e^-\else $e^+e^-$\fi}
\def\ppb{\ifmmode \bar pp\else $\bar pp$\fi}

\def\bsg{\ifmmode b\rightarrow s\gamma \else $b\rightarrow s\gamma$\fi}
 
\newskip\zatskip \zatskip=0pt plus0pt minus0pt
\def\matth{\mathsurround=0pt}

\def\atversim#1#2{\lower0.7ex\vbox{\baselineskip\zatskip\lineskip\zatskip
  \lineskiplimit 0pt\ialign{$\matth#1\hfil##\hfil$\crcr#2\crcr\sim\crcr}}}

\renewcommand{\thefootnote}{\fnsymbol{footnote}}

\begin{document} \begin{titlepage} 
\setcounter{page}{1}
\thispagestyle{empty}
\rightline{\vbox{\halign{&#\hfil\cr
&SLAC-PUB-7643\cr
&September 1997\cr}}}
\vspace{0.8in} 
\begin{center}

{\Large\bf Leptoquarks: Pride and Prejudice}
\footnote{Work supported by the Department of 
Energy, Contract DE-AC03-76SF00515}
\medskip

\normalsize THOMAS G. RIZZO
\\ \smallskip
{\it {Stanford Linear Accelerator Center\\Stanford University, 
Stanford, CA 94309}}\\ 

\end{center} 

\begin{abstract} 

Attempts to understand the recent observation of an excess of events in the 
neutral and charged current channels at high$-Q^2$ at HERA has provided an 
excellent example of how experiments at both low and high energies can be 
used to simultaneously constrain scenarios which predict new physics beyond 
the Standard Model. In this talk I will discuss this subject from the point 
of view of the construction of new models of leptoquarks. 

\end{abstract} 

\vskip0.95in
\begin{center}

Talk given at the {\it Workshop on Physics Beyond the Standard Model: Beyond 
the Desert: Accelerator and Nonaccelerator Approaches}, Ringberg Castle, 
Germany, June 8-14, 1997.

\end{center}

\renewcommand{\thefootnote}{\arabic{footnote}} \end{titlepage} 

\section{Introduction: Physics Beyond the Standard Model}

It is a widely held belief that new physics must exist beyond that predicted by 
the Standard Model(SM) if for no other reason than that it leaves us with too 
many unanswered questions and too many free parameters. Just when, where 
and how new physics will make its 
first appearance has been--and continues to be--a matter of some speculation. 
In the past few years we have seen a number of potential new physics 
signatures vanish as either more statistics was accumulated or the data and 
analysis thereof improved. These signatures have each in their turn induced 
some excitement in the community with the hope that a new window beyond the SM 
was finally opening. Though later vanishing, these affects have each taught us 
something new about what kinds of models can be constructed, even if they were 
not necessarily realized in nature, given the ever-tightening constraints 
provided by experiment. Still, new physics is out there somewhere waiting to 
be discovered; it is only a question of looking. We should learn to expect the 
unexpected. 

Searching for new physics is a multi-pronged attack on the unknown. While the 
production of a non-SM particle at a collider would be the most obvious and 
undeniable signature that one can imagine, the first sign of something new 
maybe more subtle. For example, one can imagine a significant deviation from SM 
expectations in a precision measurement, \eg, the $W$ mass or polarized 
forward-backward asymmetry for $b-$quarks, $A_b$. Instead, one might imagine 
the observation of a process forbidden by the SM, such as $\mu \to e\gamma$. 
However, as is 
well known, new physics rarely contributes to only one of these scenarios. 
For example, SUSY leads to new particle production at colliders, a potentially  
observable shift in the $W$ mass and/or $\sin^2\theta_{eff}$ and can enhance 
many rare decay processes beyond their SM expectations. The same is of course 
true with other forms of new physics. In fact, it likely that once new physics 
is found all three types of experiments will be necessary to unravel its 
detailed nature. 
Attempts to understand the excess of events at high$-Q^2$ recently observed 
at HERA in both the neutral current(NC) and charged current(CC) 
channels{\cite {hera}} provides just such an example of the strong interplay 
between the various new physics search scenarios and the constraints that 
arise from the three classes of experiments.

\section{Leptoquark Model Requirements}

If the HERA excesses are real and non-resonant 
then possible explanations include, \eg, the presence of higher dimensional 
operators{\cite {vb}} signalling compositeness or exotic modifications 
in the parton densities at large $x${\cite {parton}}; both these 
proposed scenarios face some very serious difficulties. 
Instead, if the excess is resonant a popular{\cite {big,old}} explanation 
is the $s-$channel production of a $\simeq 200-220$ GeV 
scalar(\ie, spin-0) leptoquark(LQ) with fermion number($F$) equal to zero--the 
subject of the present work. (Given the most recent results{\cite {straub}} it 
appears that the H1 excess in the NC channel is apparently 
clustered at $201\pm 5$ GeV 
while that for ZEUS is at $219\pm 9$ GeV. Whether this is reconcilable with a 
single resonance is still unknown.) How did we so quickly deduce the LQ spin 
and $F$ quantum number from the data?

Any discussion of LQ models has been historically based on the classic work 
by Buchm\" uller, R\" uckl and Wyler(BRW){\cite {brw}}. Those authors provided  
not only a set of assumptions under which consistent LQ models can be 
constructed but 
then classified them according to their possible spins and fermion number thus 
leading to the 10 states displayed in Table 1. These assumptions may be 
stated as follows: 

(a) LQ couplings must be invariant with respect to the SM gauge interactions

(b) LQ interactions must be renormalizable

(c) LQs couple to only a single generation of SM fermions

(d) LQ couplings to fermions are chiral

(e) LQ couplings separately conserve Baryon and Lepton numbers

(f) LQs only couple to the SM fermions and gauge bosons

{\small
\begin{table}
\begin{center}
\caption{Quantum numbers and fermionic coupling of the leptoquark states.  No
distinction is made between the representation and its conjugate. $B_\ell$ 
is the branching fraction of the LQ into the $ej$ final state and $Q$ is its 
electric charge.}
\begin{tabular}{l|ccrccc}
\hline\hline
 & Leptoquark \rule{0pt}{15pt} 
 & SU(5) Rep & & $Q$  & Coupling & $B_\ell$ \\ \hline
Scalars & & & & & & \\ \hline
$F=-2$ \rule{0pt}{15pt}& $S_{1L}$        
       & {\bf 5} & 
       & $1/3$ & $\lambda_L\, (e^+\bar u)$, 
$\lambda_L\, (\bar\nu\bar d)$ & $1/2$ \\
       & $S_{1R}$        
       & {\bf 5} & 
       & $1/3$ & $\lambda_R\, (e^+\bar u)$ & 1 \\
       & $\widetilde S_{1R}$ 
       & {\bf 45} & 
       & $4/3$ & $\lambda_R\, (e^+\bar d)$ & 1 \\[2ex]
       &         &        
       & & $4/3$ & $-\sqrt 2\lambda_L\, (e^+\bar d)$ & 1 \\[-4ex]
       & $S_{3L}$        
       & {\bf 45} & $\left\{\rule{0pt}{30pt}\right.$
       & $1/3$& $-\lambda_L\, (e^+\bar u)$,
$-\lambda_L\, (\bar\nu\bar d)$ & $1/2$ \\[-4ex]
       &         &        
       &         & $-2/3$& $\sqrt 2\lambda_L\, (\bar\nu\bar u)$ & 0 \\[2ex]
\raisebox{-2ex}[0pt]{$F=0$} &  
\raisebox{-2ex}[0pt]{$R_{2L}$} &        
\raisebox{-2ex}[0pt]{{\bf 45}} &
\raisebox{-2ex}[0pt]{$\left\{\rule{0pt}{20pt}\right.$} 
& $5/3$ & $\lambda_L\, (e^+u)$ & 1 \\[-4ex]
       &  &  &        & $2/3$ & $\lambda_L\, (\bar\nu u)$ & 0 \\[2ex]
       & 
\raisebox{-2ex}[0pt]{$R_{2R}$} &
\raisebox{-2ex}[0pt]{{\bf 45}} & 
\raisebox{-2ex}[0pt]{$\left\{\rule{0pt}{20pt}\right.$}
& $5/3$ & $\lambda_R\, (e^+u)$ & 1 \\[-4ex]
  & &    &  & $2/3$ & $-\lambda_R\, (e^+d)$ &  1\\[2ex]
       & 
\raisebox{-2ex}[0pt]{$\widetilde R_{2L}$} & 
\raisebox{-2ex}[0pt]{{\bf 10/15}} & 
\raisebox{-2ex}[0pt]{$\left\{\rule{0pt}{20pt}\right.$} &
$2/3$ &$\lambda_L\, (e^+d)$ & 1 \\[-4ex] 
       &       &    &         & $-1/3$& $\lambda_L\, (\bar\nu d)$ 
&  0 \\[1.2ex] \hline
Vectors & & & & & \\ \hline
\rule{0pt}{15pt}
\raisebox{-3ex}[0pt]{$F=-2$} & 
\raisebox{-2ex}[0pt]{$V_{2L}$} & 
\raisebox{-2ex}[0pt]{{\bf 24}} & 
\raisebox{-2ex}[0pt]{$\left\{\rule{0pt}{20pt}\right.$} &
 $4/3$ & $\lambda_L\, (e^+\bar d)$ & 1 \\[-4ex]
       & & && $1/3$ & $\lambda_L\, (\bar\nu\bar d)$ & 0 \\[2ex]
       & 
\raisebox{-2ex}[0pt]{$V_{2R}$} & 
\raisebox{-2ex}[0pt]{{\bf 24}} &
\raisebox{-2ex}[0pt]{$\left\{\rule{0pt}{20pt}\right.$} & 
$4/3$ & $\lambda_R\, (e^+\bar d)$ & 1 \\[-4ex]
       &   &              &         &
        $1/3$ & $\lambda_R\, (e^+\bar u)$ & 1 \\[2ex]
       & 
\raisebox{-2ex}[0pt]{$\widetilde V_{2L}$} & 
\raisebox{-2ex}[0pt]{{\bf 10/15}} &
\raisebox{-2ex}[0pt]{$\left\{\rule{0pt}{20pt}\right.$} &
 $1/3$ & $\lambda_L\, (e^+\bar u)$ & 1 \\[-4ex]
       &       &          &         &
        $-2/3$ & $\lambda_L\, (\bar\nu\bar u)$ & 0 \\[2ex]
$F=0$  & $U_{1L}$  & {\bf 10} && $2/3$ & $\lambda_L\, (e^+d)$,
$\lambda_L\, (\bar\nu u)$ & $1/2$ \\
       & 
$U_{1R}$  & {\bf 10} &&  $2/3$ & $\lambda_R\, (e^+\bar d)$ & 1 \\
       & 
$\widetilde U_{1R}$ & 
{\bf 75} &&
 $5/3$ & $\lambda_R\, (e^+u)$ & 1 \\[2ex]
       &    &             &         &
        $5/3$ & $\sqrt 2\lambda_L\, (e^+u)$ & 1 \\[-4ex]
       & 
$U_{3L}$  & 
{\bf 40} &
$\left\{\rule{0pt}{30pt}\right.$ &
 $2/3$ & $-\lambda_L\, (e^+d)$, $\lambda_L\, (\bar\nu u)$ & $1/2$\\[-4ex]
       &      &           &         &
        $-1/3$ & $\sqrt 2\lambda_L\, (\bar\nu d)$ & 0 \\[2ex]
\hline\hline
\end{tabular}
\end{center}
\end{table}}
\noindent
If we strictly adhere to these rules then the requirements of gauge invariance 
and renormalizability fix all of the spin-1 LQ couplings and thus 
its production cross section at the Tevatron{\cite {hrphp}} is simply a 
function of its mass. The possibility 
that such particles can exist in the mass range below approximately 
350 GeV can then be excluded 
based on the direct searches by both CDF and D0{\cite {cdfd0}}. (As noted by 
Bl\"umlein{\cite {big}}, the introduction of non-renormalizable anomalous 
couplings for the LQ may allow us to somewhat soften this conclusion but 
spin-1 LQs are still found to be excluded in the range of interest for the 
HERA excess.) 
If the LQ is a scalar but is of the $F=2$ type then we would have expected 
to see an event 
excess show up in the $e^-p$ NC channel and not the $e^+p$ channel as is the 
case. This is demonstrated in Table 2 which shows the event rate for each of 
the BRW scalar LQs assuming a Yukawa coupling strength of 
$\tilde \lambda=\lambda/e=0.1$ and a mass of 200 GeV normalized to a 
luminosity of 100 $pb^{-1}$. Even with the great disparity in integrated 
luminosity collected by the experiments in both channels and allowing for the 
free adjustment of the strength of the Yukawa coupling we must conclude that 
the LQ is of the $F=0$ type. We note from Table 1 that all $F=0$ scalar LQs 
must have $B_\ell=1$ and lie in $SU(2)_L$ doublets, implying that more than 
one type of LQ must exist. Given the recent strengthening of the Tevatron 
search reach and the possible CC excess at HERA, this poses a serious 
challenge to the scalar LQ interpretation, although not as serious as was 
found in the case of vector LQs. Using the next-to-leading order cross 
section formulae of Kr\"amer \etal {\cite {kramer}}, the $95\%$ 
CL lower limit on the mass of a $B_\ell=1$ scalar LQ is found by D0 to 
be 225 GeV. 
D0 has also performed a combined search for first generation leptoquarks by 
using the $eejj$, $e\nu jj$ and $\nu \nu jj$ channels. For fixed values of the 
leptoquark mass below 225 GeV, these search constraints can be used to place 
an upper limit on $B_\ell$. For $M_{LQ}$=200(210,220) GeV, D0 obtains the 
constraints $B_\ell \leq 0.45(0.62,0.84)$ at $95\%$ CL. Of course 
if CDF and D0 combine their searches in the future, then the 225 GeV bound
may rise to $\simeq 240$ GeV, in which case even 
stronger upper bounds on $B_\ell$ will be obtained. Allowing the LQ to have 
decays into the $\nu j$ final state with 
a reasonable branching fraction would solve this problem and would yield 
the desired CC signal at HERA.  
However the models in Table 1 do not allow for this possibility. 

\begin{table}
\begin{center}
\caption{Expected number of events per $100^{-1}pb$ for each electron charge 
and state of polarization for a 200 GeV scalar leptoquark at HERA assuming 
$0.4<y<1$, $\tilde \lambda=0.1$, and an electron-jet invariant mass 
$M_{ej}=200\pm 20$ GeV. These results have been smeared with a detector 
resolution of $5\%$ in $M_{ej}$.}
\begin{tabular}{lcccc}
\hline\hline
Leptoquark&$N_L^-$ & $N_R^-$ & $N_L^+$ & $N_R^+$ \\
\hline
SM background   & 51.7 & 28.7 & 9.98 & 20.0 \\
$S_{1L}$        & 121. & 28.7 & 9.98 & 20.4 \\
$S_{1R}$        & 51.7 & 167. & 10.8 & 20.0 \\
$\widetilde S_{1R}$ & 51.7 & 63.0 & 11.5 & 20.0 \\
$S_{3L}$        & 190. & 28.7 & 9.98 & 23.5 \\
$R_{2L}$        & 52.4 & 28.7 & 9.98 & 158. \\
$R_{2R}$        & 51.7 & 29.4 & 148. & 20.0 \\
$\widetilde R_{2L}$ & 53.2 & 28.7 & 9.98 & 54.4 \\
\hline\hline
\end{tabular}
\end{center}
\end{table}

How do we interpret these conflicting demands? It is clear that the BRW 
structure must be too restrictive and so conditions (a)-(f) must be 
critically re-examined. While the assumptions of gauge invariance 
and renormalizability are unquestionable requirements of LQ model building, it 
is possible that the other conditions one usually imposes are much too 
strong--unless they are specifically demanded by data. This observation 
implies that for LQs to be experimentally accessible now, or anytime soon, 
their couplings to SM fermions must be essentially purely chiral and must also 
separately conserve 
both Baryon and Lepton numbers. The condition that LQs couple to only a single 
SM generation is surely convenient by way of avoiding the numerous low energy 
flavor changing neutral current constraints{\cite {rev}} but is far from 
natural in the mass 
eigenstate basis. A short analysis indicates{\cite {old}} that the natural 
imposition of this condition in the original weak basis for the first 
generation LQ and then allowing for CKM-like 
intergenerational mixing does not obviously get us into any trouble with 
experimental constraints especially in lepton generation number is at least 
approximately conserved. However, this does not give us the flexibility we 
need to avoid the Tevatron bounds or to induce an excess in the CC channel. 
Clearly then, to obtain a new class of LQ models the LQs themselves must be 
free to couple to more than just the SM fermions and gauge fields. 
Note that assumption (f) effectively requires that the LQ be the only new 
component added to the SM particle spectrum which seems quite unlikely in any 
realistic model; this assumption must be dropped. 

What kind of LQ interaction do we want? 
In order to satisfy the HERA and Tevatron constraints it is clear that we need 
to have an $F=0$ scalar LQ as before, preferably an isosinglet so that we 
do not have several LQ states of various masses to worry about, but now with 
an effective coupling to SM fermions such as 
\begin{equation}
{\cal L}_{wanted} = [\lambda_u \nu u^c+\lambda_d ed^c]\cdot LQ +h.c.\,,
\end{equation}
with comparable values of the effective Yukawa couplings $\lambda_u$ and 
$\lambda_d$ 
thus fixing the LQ's electric charge, $Q(LQ)=\pm 2/3$. An alternative 
possibility, allowing for either Dirac neutrinos or a $\nu^c$ which is light 
and appears as missing $p_T$ in a HERA or Tevatron detector, is the interaction 
\begin{equation}
{\cal L}_{wanted}' = [\lambda_u' \nu^c u+\lambda_d' e^cd]\cdot LQ' +h.c.\,.
\end{equation}
[It is important for later analyses to note that we cannot have these two 
interactions simultaneously as we would then strongly violate assumption (d).]  
It is easy to see that in either case the LQs of any other charge 
assignment cannot simultaneously couple to both $ej$ and $\nu j$ as is 
required by the HERA and Tevatron data. Unfortunately, either of the above 
Lagrangians as they stand violate assumption 
(a) in that they are not gauge invariant with respect to 
$SU(2)_L$. This implies that the desired Yukawa couplings are only 
effective ones and must arrived at from some more fundamental theory. Even if 
we are successful in obtaining one or both of these Lagrangians, is it clear 
that we can find values of $\lambda_u$ and $\lambda_d$ which are compatible 
with all of the data?

\section{Constraining Leptoquark Couplings}

What are the existing constraints on LQ Yukawa couplings? 
If the LQ has only the couplings described by one of the above Lagrangians  
then we can, \eg, trade in $\lambda_u$ for 
$B_\ell=\lambda_d^2/(\lambda_d^2+\lambda_u^2)$, since we are assuming 
that the LQ has no other decay modes. As discussed above, the 
Tevatron searches place a $\lambda_d$-independent constraint on $B_\ell$ 
for any fixed value of the LQ mass. 
Similarly, as promised, low energy measurements play an important role here as 
well. The recent constraints on the size of any allowed deviation of the weak 
charge from its SM value in Atomic Parity Violation(APV) in 
Cesium{\cite {wood}}, $\Delta Q_W=1.09\pm 0.93$, places $B_\ell$-independent 
bounds on $\lambda_d${\cite {ros}} for fixed $M_{LQ}$. Similarly, $\mu-e$ 
universality in $\pi$ decay, expressed through the ratio 
$R=\Gamma(\pi \to e\nu)/\Gamma(\pi \to \mu \nu)=0.9966\pm 0.0030$, 
constrains the product of couplings $\lambda_u \lambda_d${\cite {pdg}}.
The observed rate of NC events at HERA itself essentially constrains instead 
the product $\lambda_d^2B_\ell$; in 
the later case QCD and efficiency corrections are quite important{\cite {ks}}. 
Putting all of these together defines an approximate allowed 
region in the $B_\ell-\tilde \lambda_d$ plane shown in Fig.1 for different 
values of $M_{LQ}$. Here, we define $\tilde \lambda=\lambda/e$, with $e$ the 
conventional proton charge. (This scaling of the coupling to $e$ follows 
earlier tradition{\cite {phyrep}}.)  We note that these allowed regions are 
compatible with the cross section required to explain the HERA CC excess. 

\begin{figure}[htbp]
\centerline{
\psfig{figure=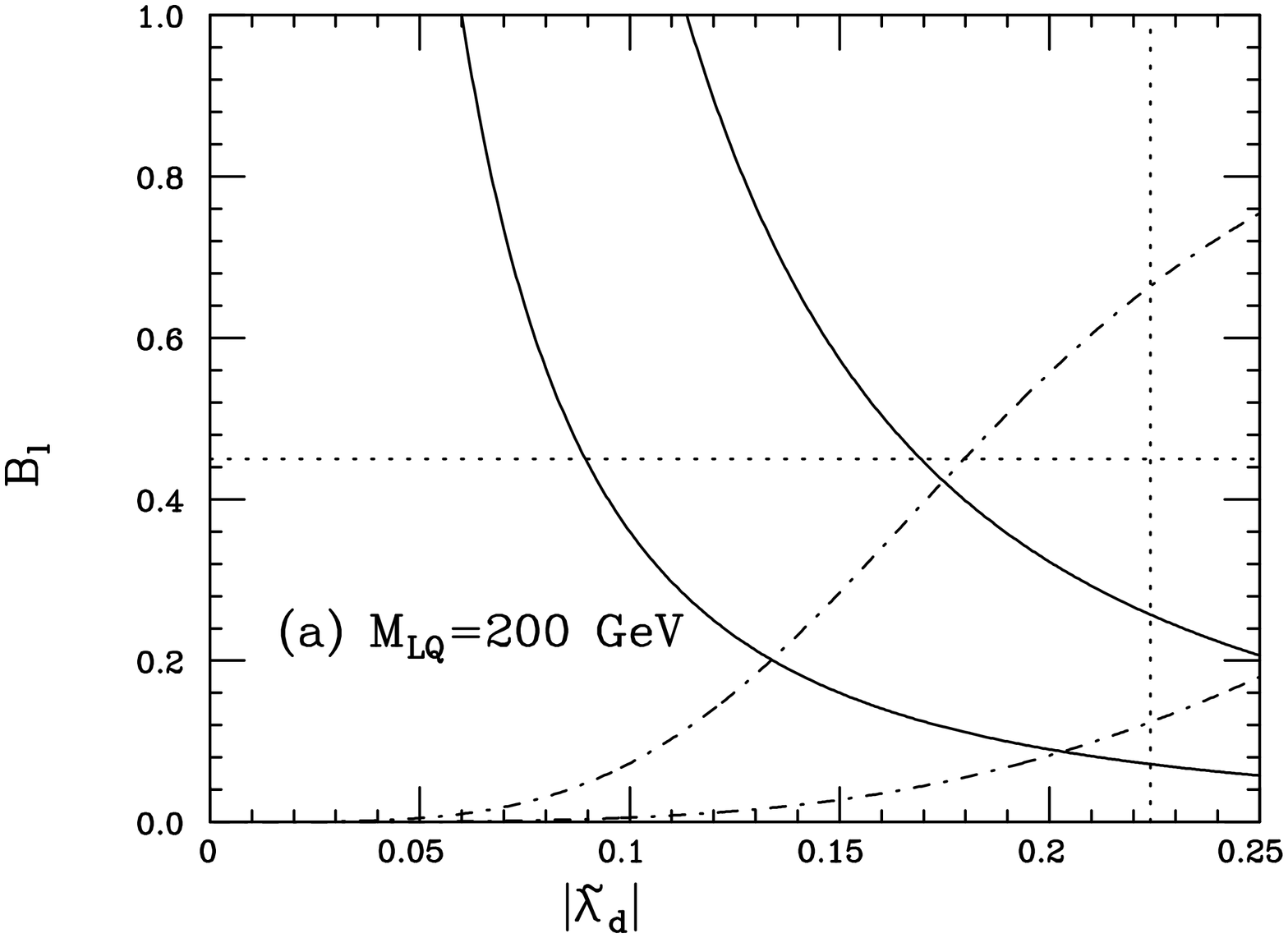,height=9.1cm,width=9.1cm,angle=-90}
\hspace*{-5mm}
\psfig{figure=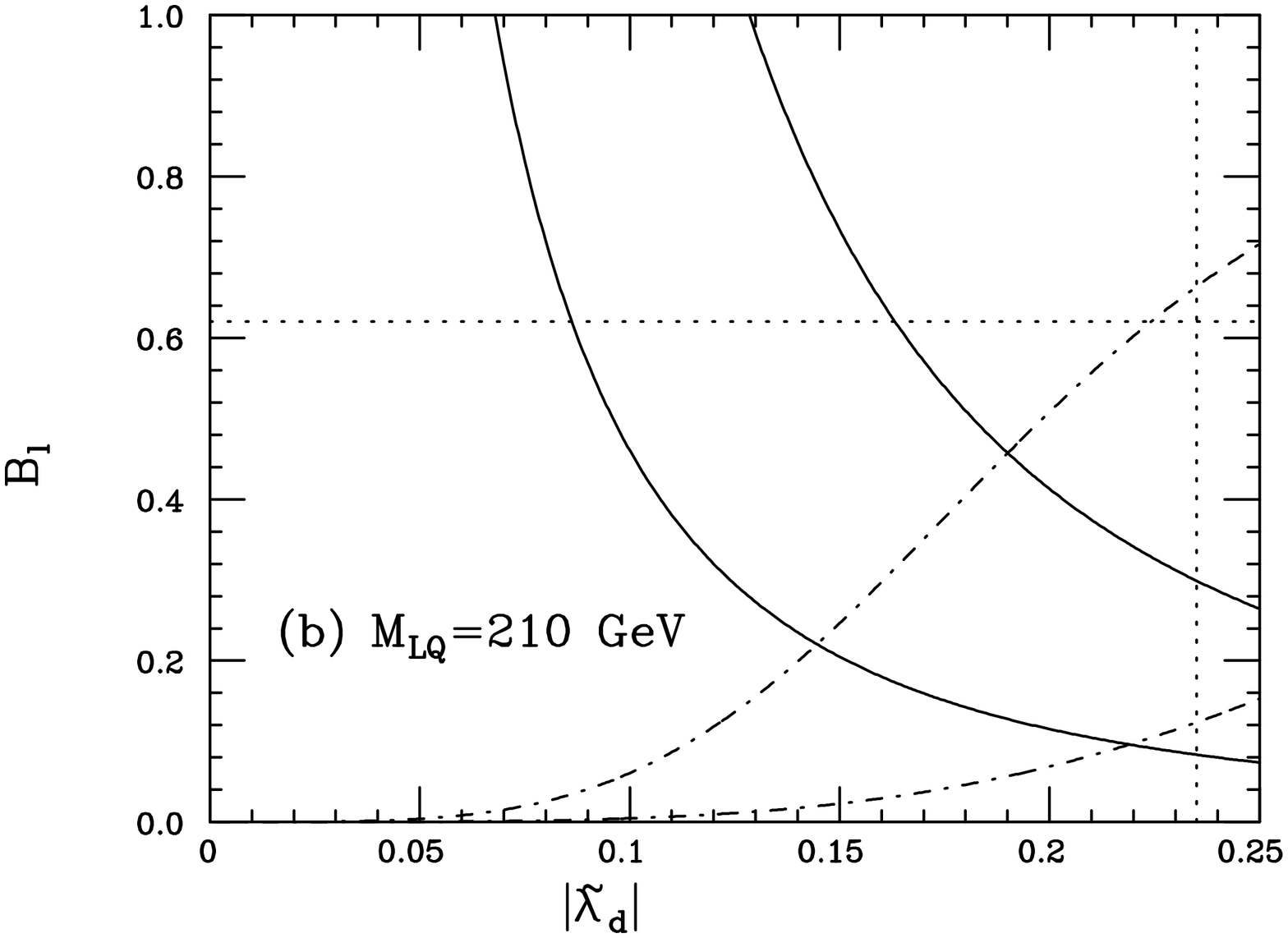,height=9.1cm,width=9.1cm,angle=-90}}
\vspace*{-0.75cm}
\centerline{
\psfig{figure=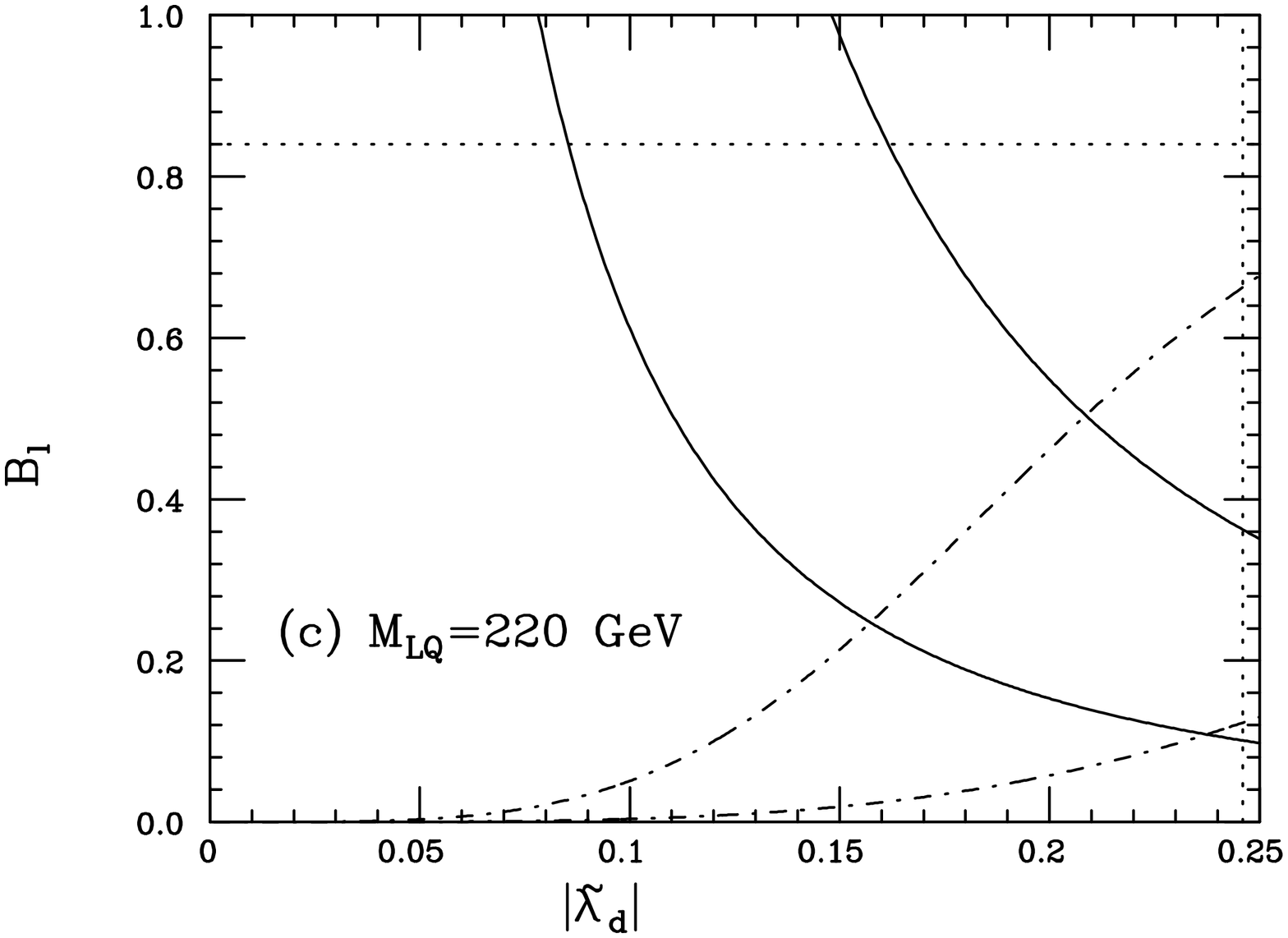,height=9.1cm,width=9.1cm,angle=-90}}
\vspace*{-1cm}
\caption{Allowed parameter space region in the $B_\ell-\tilde \lambda_d$ plane 
for a LQ with mass 200 GeV(top left), 210 GeV(top right) or 220 GeV(bottom). 
The region allowed by the direct Tevatron searches is below the horizontal 
dotted line while that allowed by APV data in Cesium is to the left of the 
vertical dotted line. 
The region inside the solid band is required to explain the HERA excess in the 
NC channel. The region above the dash-dotted curve is allowed by 
$\pi$ decay universality: the lower(upper) curve corresponds to the case 
where $\lambda_u \lambda_d >(<)0$.}
\end{figure}

There are other means to probe LQ couplings but they are somewhat more 
indirect. There are several ways in which LQs may make their presence known 
in $e^+e^-$ collisions\cite{old,us}.  At center of mass energies below the 
threshold for pair production, the existence of LQs can lead to deviations in 
both the cross section and angular distributions for $e^+e^- \to q\bar q$. This 
may be particularly relevant when $\sqrt s$ is comparable to the leptoquark 
mass as would be the case at LEP II if a 200-220 GeV LQ did exist. The origin 
of these modifications is due to the $t-$channel LQ exchange and is thus 
proportional in amplitude to the square of the unknown Yukawa coupling. 
However, in {\cite {old}} it was shown that even with large data samples it is 
unlikely that LEP II will have the required sensitivity to probe couplings 
as small as those suggested by the HERA data as shown in Fig.2. 
The OPAL Collaboration{\cite {opal}} has recently performed this analysis with 
real data at somewhat lower energies but with comparable results. 

\begin{figure}[htbp]
\centerline{
\psfig{figure=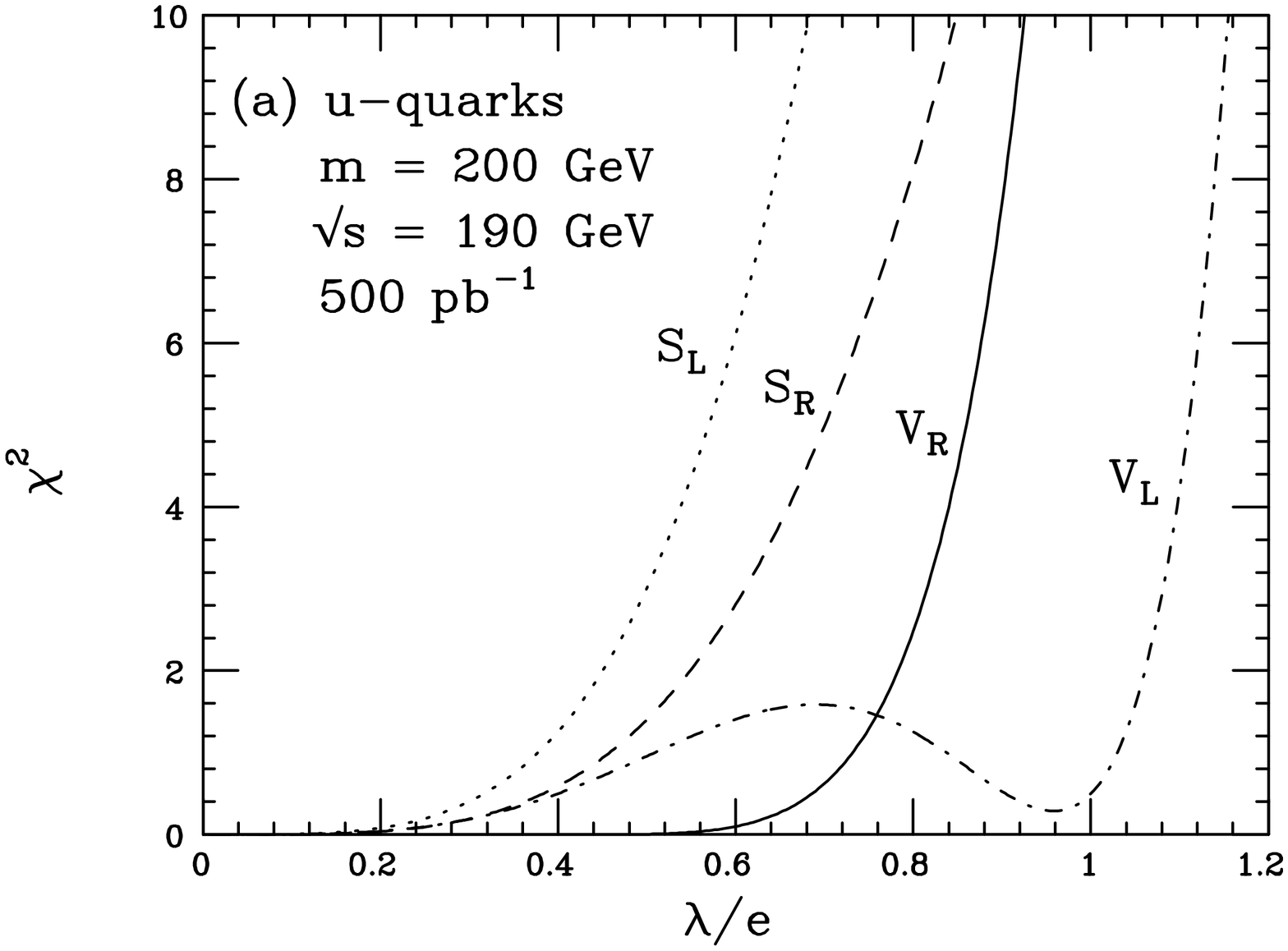,height=10.cm,width=12.cm,angle=-90}}
\vspace*{-.75cm}
\centerline{
\psfig{figure=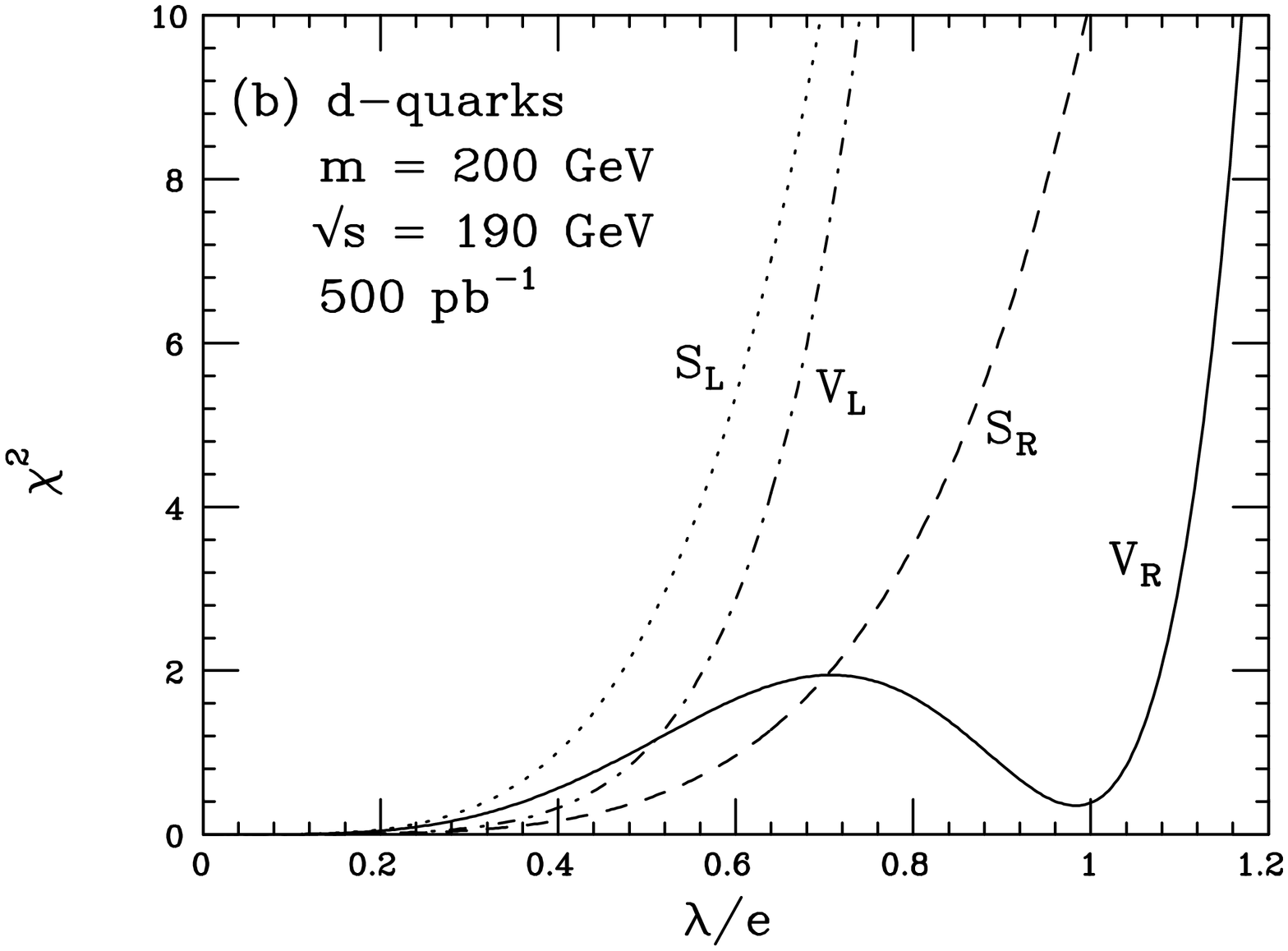,height=10.cm,width=12.cm,angle=-90}}
\vspace*{-0.6cm}
\caption{$\chi^2$ fits to the SM angular distribution for $e^+e^-\to q\bar q$ 
at 190 GeV including the effects of a 200 GeV LQ coupling to (a) u-
or (b) d-quarks. S or V labels spin-0 or spin-1 type LQs. In both cases the 
dotted(dashed) curve corresponds to a scalar 
LQ with a left(right)-handed coupling while the dash-dotted(solid) 
curve corresponds to the vector LQ case with left(right)-handed couplings.}
\end{figure}

Turning this process around, we can imagine that the Drell-Yan production of 
either $e^+e^-$ or $e^\pm \nu$ channel at the Tevatron Main Injector 
may show some 
sensitivity{\cite {dryan}} to LQ exchange in the $t-$channel. In the $e^+e^-$ 
case the observables are the invariant mass distribution and the 
forward-backward asymmetry. In the $e^\pm \nu$ channel, the corresponding 
observables are the transverse mass distribution on the electron rapidity 
asymmetry. Figs. 3 and 4 show the result of these considerations; neither 
channel has the sensitivity to probe Yukawas in the desired range if only 
2 $fb^{-1}$ of luminosity is available. 

\vspace*{-0.5cm}
\nn
\begin{figure}[htbp]
\centerline{
\psfig{figure=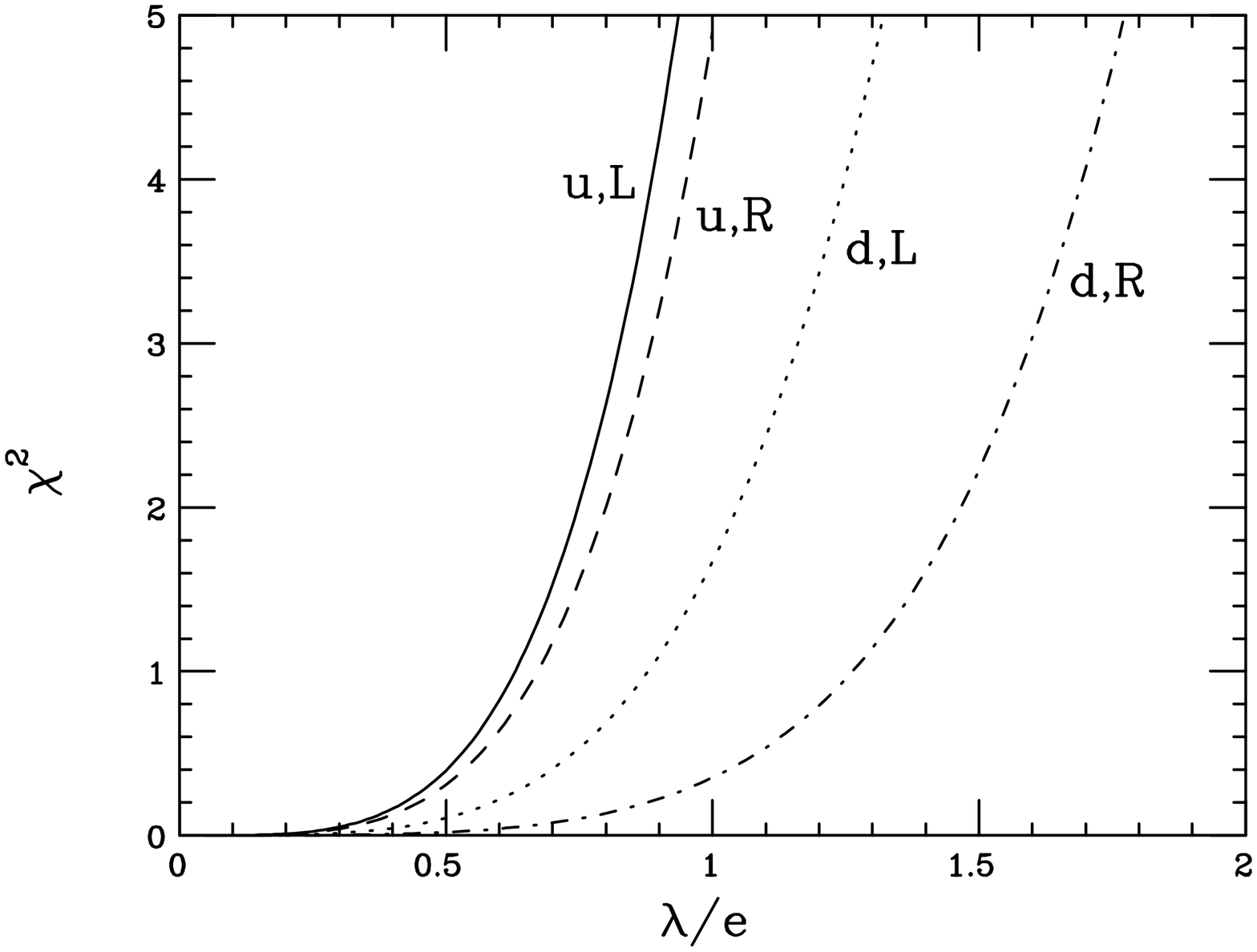,height=14.cm,width=16cm,angle=-90}}
\vspace*{-0.75cm}
\caption{$\chi^2$ fits to Drell-Yan production at the Tevatron Main Injector 
assuming a luminosity of 2 $fb^{-1}$ including the effects of a
200 GeV scalar leptoquark for each type of leptoquark coupling as labeled.}
\end{figure}

Leptoquarks can also be produced singly at hadron colliders through their 
Yukawa couplings. Compared to pair production, this mechanism has the 
advantage of a larger amount of available phase space, but has the 
disadvantage in that it is directly proportional to the small Yukawa coupling. 
For a general 200 GeV scalar leptoquark with a coupling strength of 
$\tilde \lambda=0.15$ calculations show that we would obtain 
approximately $\sim 88\,, 196$ events from 
$gd+g\bar d\,, gu+g\bar u$ fusion,  respectively, assuming $10$ 
fb$^{-1}$ of integrated luminosity at the Main Injector/TeV33.  This event 
rate should be marginally sufficient to provide a very rough determination 
of the value of Yukawa coupling $\tilde \lambda_d$ for the models of specific 
interest. 

Additional LQ coupling 
information can possibly be obtained from the sum of the squares of the first 
row of the CKM matrix, $\sum_i|V_{ui}|^2$. This involves combining experiments 
at low, medium and high energies and it an essential test of quark-lepton 
universality and CKM unitarity. In the SM this sum is, of course, unity, but 
LQ exchange can yield either an apparent upward or downward shift in the 
extracted value of $|V_{ud}|$:
\begin{equation}
|V_{ud}|^2_{eff}\simeq |V_{ud}|^2_{true}-1.52\times 10^{-3}\left( {200~GeV
\over M_{LQ} }\right)^2\left( {\tilde\lambda_u\over 0.15}\right)
\left( {\tilde\lambda_d\over 0.15}\right) \,,
\end{equation}
so that it would appear experimentally as if a unitarity violation 
were occurring. Interestingly, 
the value of the above sum has recently been discussed by Buras\cite{ajb},
who reports $\sum_i|V_{ui}|^2=0.9972\pm 0.0013$, which is more than $2\sigma$
below the SM expectation.  Clearly, if $\tilde\lambda_u\tilde\lambda_d>0$,
the LQ exchange provides one possible additional contribution which, for
$\tilde\lambda_u=\tilde\lambda_d=0.15$ and $M_{LQ}=200$ GeV,
would increase the sum to $0.9987$, now only $1\sigma$ low. This ``same sign'' 
possibility is clearly preferred by the combined set of present data as shown 
in Fig.2. This situation requires watching in the future.

\vspace*{-0.5cm}
\nn
\begin{figure}[htbp]
\centerline{
\psfig{figure=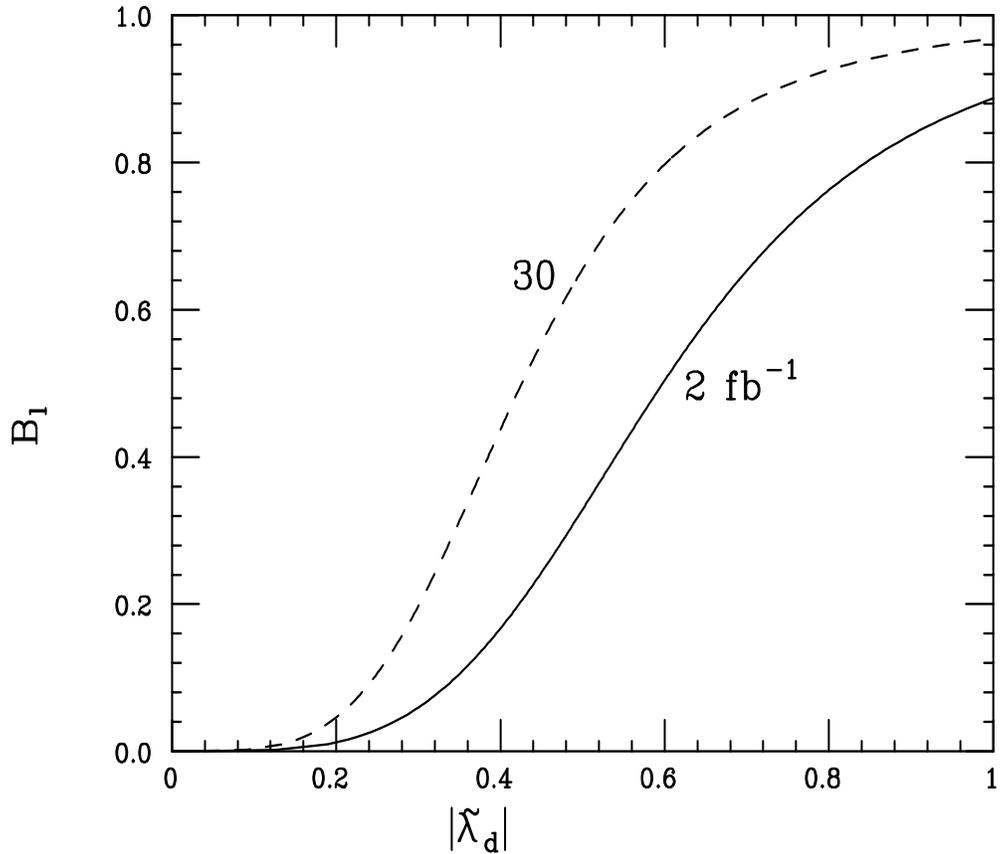,height=14cm,width=16cm,angle=-90}}
\vspace*{-0.9cm}
\caption{$95\%$ CL bound on $B_\ell$ as a function of $\tilde \lambda_d$ 
from a fit to both the $M_T$ distribution and lepton rapidity asymmetry, 
$A(\eta_\ell)$, at the Tevatron 
Main Injector for two integrated luminosities as indicated. The area below 
and to the right of the curves are excluded, a LQ of mass 200 GeV is assumed.}
\end{figure}

A last possibility is the observation of LQ exchange through radiative 
corrections. Unfortunately, it has been shown{\cite {old}} that LQs rapidly 
decouple and give only tiny contributions to the oblique 
parameters{\cite {obl}}. In addition it has also been shown that LQ exchange 
does not significantly modify $Z$-pole physics{\cite {oscar}} through vertex 
corrections.

\section{New Leptoquark Model Building}

We now turn to dealing with the construction of models that lead to one of the 
Lagrangians above. Given the fixed gauge structure of the SM the most likely 
new interactions that LQs may possess are with the Higgs field(s) responsible 
for spontaneous symmetry breaking and with new vector-like fermions that are 
a common feature in many extensions of the SM.
In this lecture we will consider and classify all models wherein heavy 
vector-like fermions(VLFs) are used to generate the effective interactions 
${\cal L}_{wanted}$ or ${\cal L}_{wanted}'$ at low energies. 
The emphasis of our approach will be to keep the VLFs as 
active participants in our models and not auxiliary devices to produce the 
desired coupling structure. We will assume that the fundamental LQ coupling is 
that between a VLF and a SM fermion. 
a LQ couples to both a VLF and that ordinary SSB induces a a mixing between the 
two sets of fermions. SSB is thus the true source of the desired LQ 
interactions and explains how effective interactions can arise that are not 
obviously gauge invariant. 
The small size of the effective Yukawa couplings in the above Lagrangians, 
${\cal L}_{wanted}$ or ${\cal L}_{wanted}'$, 
will then be subsequently explained by the same mechanism that produces the 
ordinary-exotic fermion mixing and sets the scale of the VLF masses in the 
TeV region. We note that the use of VLFs in 
this role is particularly suitable since in their unmixed state they make 
essentially no contribution to the oblique parameters{\cite {obl}}, they are 
automatically anomaly free and they can have bare mass terms which are SM 
gauge invariant. (Alternatively their masses can be generated by the vacuum 
expectation value of a SM singlet Higgs field.)

To proceed with the analysis, we first construct the six `skeleton' models that 
are obtainable by simply coupling one of the usual SM fermion representations,  
$(L,Q,u^c,d^c,e^c,\nu^c)$, with an appropriate VLF, $X_i$(or $X_i^c$), and 
the LQ field. [We use $X_i(X_i^c)$ to denote the VLF fields with $F>(<)0$.] 
Note that we have allowed for the possibility if right-handed neutrinos. 
To this we must add the bare mass term for $X_i$ as well as any gauge 
invariant terms that can be constructed using the remaining SM fermion fields, 
$X_i^c$(or $X_i$), and the SM Higgs doublet fields $H$ and $H^c$. In each case 
gauge invariance tells us the the quantum numbers of the VLFs under the 
assumption that they are either in singlets or in fundamental representations 
under $SU(2)_L$ and $SU(3)_C$. We strictly adhere to BRW constraints (a)-(e) 
in forming our constructions. These six `skeletons' are: 
\begin{eqnarray}
{\cal L}_A & = & \lambda_A LX_1^c\cdot LQ+a_uX_1u^cH+a_dX_1d^cH^c-M_1X_1X_1^c
-M_Q'QX_1^c\,, \nonumber\\
{\cal L}_B & = & \lambda_B QX_2^c\cdot LQ+a_eX_2e^cH^c+a_{\nu} X_2 \nu^cH
-M_2X_2X_2^c-M_L'LX_2^c\,, \nonumber\\
{\cal L}_C & = & \lambda_C X_3u^c\cdot LQ+a_1LX_3^cH 
-M_3X_3X_3^c-M_N'N\nu^c\,, \nonumber\\
{\cal L}_D & = & \lambda_D X_4d^c\cdot LQ+a_2LX_4^cH^c 
-M_4X_4X_4^c-M_E'Ee^c\,, \\
{\cal L}_E & = & \lambda_E X_5e^c\cdot LQ+a_3QX_5^cH^c
-M_5X_5X_5^c-M_D'Dd^c\,, \nonumber\\
{\cal L}_F & = & \lambda_F X_6\nu^c\cdot LQ+a_4QX_6^cH
-M_6X_6X_6^c-M_U'Uu^c\,, \nonumber
\end{eqnarray}
We will assume that all of the Yukawa couplings that appear in these 
`skeletons' are of order unity and that no fine-tuning is present. 
These constructs have a number of obviously desirable features but they do not 
yet have all the necessary ingredients. For example, the LQ in `skeleton' 
C(D) only couples to the $u(d)$ quark while that in `skeleton' E(F) only 
couples to $e(\nu)$. For `skeletons' A and B we see that the LQs couple to 
both $u,d$ and $e,\nu$, respectively. The solution to this problem is to 
combine the various `skeletons' into full models that have all of the desired 
couplings. This procedure is straightforward but when doing so we must take 
care not to 
violate the assumption that the LQ couplings are chiral. Given this very strong 
constraint, the entire list of models that can be constructed in this fashion 
are only ten in number: A, B, CD, EF, AC, AD, ACD, BE, BF and BEF. The 
combination of letters symbolizes that we add the respective Lagrangians and 
identify the LQ field as common. We note 
that models A, CD, AC, AD and ACD produce the interaction ${\cal L}_{wanted}$, 
while models B, EF, BE, BF and BEF produce instead ${\cal L}_{wanted}'$. As 
we will see below in each case the color, isospin and electric charge quantum 
numbers are completely fixed by gauge invariance and the assumption that the 
LQ is a $|Q|=2/3$ isosinglet with $F=0$. It is very important to remember 
that the five models leading to 
${\cal L}_{wanted}'$ would be {\it excluded} if the neutrino is not a Dirac 
field of if $\nu^c$ cannot appear as missing energy or $p_T$ in a detector. 

Having said all this we have yet to explicitly see how even one of these 
models works so we now examine model A in detail. 
Here, we have coupled an exotic fermion, denoted as $X_1$, 
to $L$ plus a leptoquark. In this case gauge invariance 
requires that $X_1$ be an isodoublet, with member charges of $2/3,-1/3$ since 
the leptoquark charge and fermion number are fixed, as well as an 
$SU(3)_C$ triplet.  
We can thus write $X_1^T=(U^0,D^0)$, where the superscript denotes
the  weak eigenstate fields. When $H$ and $H^c$ receive vevs (which we 
denote by $v$ and $v^c$, respectively), the $a_{u,d}$ terms in the
above Lagrangian induce off-diagonal couplings in both the $Q=-1/3$ and $Q=2/3$
quark mass matrices.  Neglecting the $u$- and $d$-quark masses, these are 
given in the $\bar \psi_L^0 M \psi_R^0$ weak eigenstate basis by
\begin{eqnarray}
\bar \psi_L^0 M_{u}\psi_R^0 & = & (\bar u^0,\bar U^0)_L \left( \begin{array}{cc}
0 & -M_Q'  \\
a_uv  & -M_1  
\end{array} \right)\left( \begin{array}{c}
u^0 \\
U^0
\end{array} \right)_R  \,, \\
\bar \psi_L^0 M_{d}\psi_R^0 & = &(\bar d^0,\bar D^0)_L  \left( \begin{array}{cc}
0 & -M_Q'  \\
a_dv^c  & -M_1  
\end{array} \right) \left( \begin{array}{c}
d^0 \\
D^0
\end{array} \right)_R  \,.
\end{eqnarray}
Both $M_{u,d}$ can be diagonalized by a bi-unitary transformation which 
becomes simply bi-orthogonal under the assumption that 
the elements of $M_{u,d}$ are real, resulting in the diagonal mass matrices
$M_{u,d}^{diag}=U_L(u,d)M_{u,d}U_R(u,d)^\dagger$. Since $U_{L,R}(u,d)$ are 
simple $2\times 2$ rotations they can each be 
parameterized by a single angle $\theta_{L,R}^{u,d}$. Assuming that both 
$M_1,M_Q'$ are large in comparison to either $a_uv$ or $a_dv^c$ we find 
$\theta_R^{u,d}\simeq a_{u,d}v(v^c)M_1/(M_1^2+M_Q'^2)$. With $M_1,M_Q'$ of 
order 1 TeV, $v,v^c$ of order 100 GeV and $a_{u,d}$ of order unity this 
implies that $\theta_R^{u,d}\simeq 0.05$. 
Writing $U^0\simeq U+\theta_R^u u$ in terms of the
mass eigenstate fields, and similarly for $D^0$, the interaction involving the 
SM fermions and the leptoquark thus becomes
\begin{equation}
{\cal L}_{light} = \left[\left({\lambda_A a_u vM_1\over (M_1^2+M_Q'^2)}
\right)~\nu u^c+\left({\lambda_A a_dv^cM_1\over (M_1^2+M_Q'^2)}\right)
~ed^c\right]\cdot LQ +h.c.\,,
\end{equation}
which is the exact form we desired in Eqn. (1). For $\lambda_A$ again of order 
unity this naturally leads to a 
reasonable relative branching fraction for the $LQ\to \nu j$ decay mode, and 
gives acceptable values for $\lambda_{u,d}$ in Eqn. (1) for $M_1,M_Q'$ in the 
TeV range. Note that $\theta_R^{u,d}\neq 0$ leads to a modification of both 
the $u$ and $d$ quark couplings to the $Z$ and induces $W$-mediated 
right-handed charged current interactions as well.  

What about $\theta_L^{u,d}$? $\theta_L^{u,d}\neq 0$ does not contribute to 
the LQ couplings or influence $Z$ couplings of $u,d$ since the left-handed 
SM fermions and the VLFs have the same quantum numbers. With $M_1$ and $M_Q'$ 
of comparable size both $\theta_L^{u,d}$ are found to be large and of almost 
identical magnitude. However, modifications to the left-handed CC couplings 
of $u$ and $d$ to the $W$ are only 
sensitive to the deviation $\Delta=1-\cos(\theta_L^u-\theta_L^d)$. Here, for 
$M_1=M_Q'$, the difference 
$\theta_L^u-\theta_L^d\approx (a_u^2v^2-a_d^2v^{c2})/4M_1^2$ is found to be 
very small, of order $\sim \theta_R^2\simeq (0.05)^2$. This implies that 
$\Delta$ itself is of only order $10^{-5}$ or less--practically invisible. 

The other models above work more or less in a similar fashion except that in 
most cases mixing is taking place between a number of different SM fermions and 
their VLF partners. The effective couplings in Eqns. 1 and 2 then derive from 
more than a single source. For completeness Table 3 identifies the VLFs which 
are present in each of these models. 
\begin{table}
\centering
\begin{tabular}{|c|c|} \hline\hline
Model & Vector-like Fermions \\ \hline
 & \\
A & $\left( \begin{array}{c}
            U \\
            D
\end{array} \right)_{L,R}$ \\[4ex]
CD & $N_{L,R}; E_{L,R}$ \\[2ex]
AC & $\left( \begin{array}{c}
            U \\
            D
\end{array} \right)_{L,R}; N_{L,R}$ \\[4ex]
AD & $\left( \begin{array}{c}
            U \\
            D
\end{array} \right)_{L,R}; E_{L,R}$ \\[4ex]
ACD & $\left( \begin{array}{c}
            U \\
            D
\end{array} \right)_{L,R}; N_{L,R}; E_{L,R}$ \\[4ex]
\hline
 & \\
B & $\left( \begin{array}{c}
            N \\
            E
\end{array} \right)_{L,R}$ \\[4ex]
EF & $U_{L,R}; D_{L,R}$ \\[2ex]
BE & $\left( \begin{array}{c}
            N \\
            E
\end{array} \right)_{L,R}; D_{L,R}$ \\[4ex]
BF & $\left( \begin{array}{c}
            N \\
            E
\end{array} \right)_{L,R}; U_{L,R}$ \\[4ex]
BEF & $\left( \begin{array}{c}
            N \\
            E
\end{array} \right)_{L,R}; U_{L,R}; D_{L,R}$ \\[4ex]
\hline\hline
\end{tabular}
\caption{Listing of models and the new vector-like fermions which
are contained in them.}
\label{vlf}
\end{table}

\section{Tests}

To directly test the proposed models we must look for new physics signatures 
{\it beyond} those suggested by the LQ interactions described Eqns. 1 and 2. 
One possibility is to probe for the additional interactions between the VLFs 
and the LQ; this is obviously difficult since the VLFs are so massive. A 
second possibility is to directly look for the influences of the VLFs 
themselves. For the time being this must be an indirect search since the LHC 
will be required to directly produce the VLFs in the range of interest to us 
here. Once the LHC is available, however, a 1 TeV color triplet will produce 
1000 events/yr even with a luminosity of 10 $fb^{-1}$. 

In the first category a potential new process of interest is the pair 
production of like sign LQs at the Tevatron through $t-$ and $u-$channel 
$N$ exchange (in those models where it is present) when $N$ is a Majorana 
field: $uu\to 2LQ$. This rate for this process goes as the fourth power of the  
$\lambda_B$ or $\lambda_C$ Yukawa coupling but these are assumed to be of 
order unity so that potentially large cross sections are obtainable. 
One finds the subprocess cross section to be
\begin{equation}
{d\sigma\over {d\hat t}} ={\lambda^4\over {64\pi \hat s}} 
\left[{M_N(\hat t+\hat u-2M_N^2)
\over {(\hat t-M_N^2)(\hat u-M_N^2)}}\right]^2\,,
\end{equation}
where $M_N$ is the mass of the $N$. Note that as $M_N \to 0$ the rate 
vanishes as one might expect for a Majorana fermion induced process. The 
cross section for this reaction at the Tevatron Main Injector for $\lambda=1$ 
and $M_{LQ}=200$ GeV is approximately 50 fb for $M_N$=1 TeV and falls off 
quickly with increasing as $M_N$ increases. Since the signature for this 
process is 2 jets plus like-sign leptons there is little SM backgrounds and 
so it may be observable during Run II. 

The best indirect tests for the presence of VLFs which mix with the 
conventional fermions are searches for new physics associated with deviations 
in couplings from SM expectations{\cite {phyrep,dpf}}. Two of the best tests 
here are our old friends quark-lepton universality in the guise of $V_{ud}$ 
and the leptonic decays 
of the $\pi$ discussed above. To clarify this point we note that in models 
where, \eg, the $u$ and $d$ mix with isodoublet VLFs the $W$ couples as 
$\bar ud_L\cos(\theta_L^u-\theta_L^d)+\bar ud_R\sin \theta_R^u\sin \theta_R^d$ 
whereas if the VLFs are isosinglets the corresponding coupling is 
$\bar ud_L\cos \theta_L^u\cos \theta_L^d$. (The corresponding couplings can be 
also written down for the case of leptonic mixing.) We note also the general 
feature we find 
is that in models where the VLFs are in isodoublets $\theta_R's \sim 0.05$ 
and differences in $\theta_L's \sim (0.05)^2$, whereas the converse is true 
when the relevant VLFs are isosinglets. 

Leptonic mixing will never show up as a shift in the value of $|V_ud|^2$ since 
the modification in the amplitude occurs not only in the process 
$n\to pe\bar \nu$ but also in $\mu$ decay so that it is absorbed into the 
definition of $G_F$. On the otherhand if quark mixing occurs $|V_ud|^2$ will 
experience an apparent small shift(to leading order in the mixing angles) 
$\sim -(\theta_L^u-\theta_L^d)^2+(\theta_R^u \theta_R^d)^2$ in models with 
isodoublet VLFs and $\sim -(\theta_L^u)^2-(\theta_L^d)^2$ in the isosinglet 
VLF case. For the ratio $R$ in $\pi$ decay any modification of the hadronic 
matrix element will factor out so that there is no sensitivity to quark mixing. 
However, leptonic mixing no longer factorizes and we find a shift in $R$ by 
an amount $\sim  -(\theta_L^\nu-\theta_L^e)^2-(\theta_R^\nu \theta_R^e)^2$ 
in the case of isodoublet mixing and $\sim -(\theta_L^\nu)^2-(\theta_L^e)^2$ 
for isosinglets. Recall that isodoublet leptonic mixing is only viable in 
models where the neutrino is Dirac or the right-handed neutrino appears as an 
ordinary neutrino. In these same isodoublet models, the presence of both 
left- and right-handed CC couplings and heavy VLFs can lead to a contribution 
to the $g-2$ of the electron and $\nu$ which are typically both or order 
$a ~few \cdot 10^{-11}$, neither of which are far from the present level of 
sensitivity. 

The mixing of the SM fermions with the VLF modify their couplings to the $Z$. 
These are difficult to observe particularly in the case of quarks due to the 
small size of the effect and QCD correction uncertainties. In the case of 
leptonic mixing there is not only the contribution due to mixing but there is 
an overall normalization change in the couplings due to our redefinition of 
$G_F$. This mixing can lead to a shift in $Z\to e^+e^-$ width by $\simeq 0.2$ 
MeV and an apparent shift in $\sin ^2 \theta_{eff}$ from the asymmetries of 
$\simeq 0.0006$. Again, shifts of this size are near the present limit of 
experimental sensitivity. In a similar manner SM expectations for $Q_W$ in 
APV measurements may also be modified if SM mixing with the VLF occurs. 
However, in this case it is easy to show that the fractional change in $Q_W$ 
due to these effects is only at the level of $\sim 10^{-3}$.

\section{Grand Unification with Leptoquarks}

If LQs are indeed real and we also believe that there is experimental 
evidence for coupling constant unification then we must begin to examine 
schemes which contain both ingredients as pointed out in{\cite {old}}. In the 
scenarios presented here the SM quantum numbers of the LQ are fixed 
but new VLFs have now been introduced as well, all of which 
will alter the usual RGE analysis of the running couplings. 

Before discussing SUSY models we note with some curiosity that coupling 
unification {\it can} occur in LQ models containing exotic fermions 
even if SUSY is not introduced as was shown many years ago in 
{\cite {my,oldt}}. Of 
course in the work of Murayama and Yanagida {\cite {my}}, the LQ was an 
isodoublet and one of the particular models on the BRW list, now excluded by 
the combined HERA and Tevatron data. In the scenarios presented above the 
LQ is now a $Q=2/3$ isosinglet so that the Murayama and Yanagida 
analysis does immediately apply. Fortunately, we see from the results of 
Ref. {\cite {oldt}} that a second possibility does exist for just this case: 
one adds to the SM spectrum the LQ and its conjugate as well as a 
vector-like pair of color-triplet, 
isodoublets together with the field $H^c$. This is the just particle 
content of the model A. To verify and update this analysis, let us assume 
for simplicity that all the new matter fields are introduced at the weak 
scale and take $\sin ^2 \theta_w=0.2315$ as input to a two-loop RGE analysis. 
We then obtain the 
predictions that coupling unification occurs at $3.5 \times 10^{15}$ GeV and 
$\alpha_s(M_Z)$ is predicted to be 0.118. If unification does indeed occur we 
can estimate the proton lifetime{\cite {mrbill}} to be 
$\tau_p=1.6\times 10^{34\pm 1}$ years, safely above current 
constraints{\cite {pdg}}. We find this situation to be rather intriguing and we 
leave it to the reader to further ponder. 

Of course there are other reasons to introduce SUSY beyond that of 
coupling constant unification. This subject has been discussed at some length 
in {\cite {old}} from which we extract several important observations:
($i$) To {\it trivially} preserve the 
successful unification of the SUSY-SM, only complete $SU(5)$ representations 
can be added to the MSSM spectrum. As is well-known, the 
addition of extra matter superfields in complete $SU(5)$ representations 
delays unification 
and brings the GUT scale closer to the string scale. Of course, 
there still remains the rather unnatural possibility of adding incomplete, but 
`wisely chosen', split representations. Employing split representations 
certainly allows for more 
flexibility at the price of naturalness but still requires us to choose 
sets of $SU(3)_C\times SU(2)_L \times U(1)_Y$ representations 
which will maintain asymptotic freedom and perturbative unification. 
An example of this rather bizarre scenario is the possibility of 
adding a $(2,3)(1/6)$ from a {\bf 15} and a
$(1,1)(1)\oplus (1,\bar 3)(-2/3)$ from a {\bf 10} to the low energy 
spectrum{\cite {old}}. Here the notation 
refers to the $(SU(3)_C,SU(2)_L)(Y/2)$ quantum numbers of the representation. 
We remind the reader that the LQ itself transforms as 
$(1,3)(2/3)$; the smallest standard $SU(5)$ representation into which the 
${\mbox{LQ}}+{\mbox{LQ}}^c$  
can be embedded is a {\bf 10}$\oplus\overline{\mbox{\bf 10}}$ while in 
flipped-$SU(5)\times U(1)${\cite {flip}}, it can be placed in a 
{\bf 5}$\oplus\overline{\mbox{\bf 5}}$. 
($ii$) Since we are using VLFs in these models, it is clear that 
only pairs of representations,  {\bf R}$+\overline{\mbox{\bf R}}$, can be 
added to the MSSM spectrum in order to maintain anomaly cancelation. Of 
course this is also true for the LQ superfield in that 
both LQ and $LQ^c$ fields must now be present as discussed above. 
($iii$) To preserve perturbation 
theory and asymptotic freedom up to the GUT scale when adding complete 
representations, at most one {\bf 10}$+\overline{\mbox{\bf 10}}$ or 
three {\bf 5}$+\overline{\mbox{\bf 5}}$ can be appended to the low energy 
spectrum of the MSSM apart from SM singlets. The reason for this is the 
general observation that if one adds more than 
three, vector-like, color triplet superfields to the MSSM particle content then 
the one-loop QCD beta function changes sign. Recall that the LQ itself 
{\it already} accounts for one of these color triplets. This same 
consideration also excludes the introduction of light exotic fields in 
higher dimensional $SU(3)_C$ representations. Complete $SU(5)$ 
representations larger than {\bf 10}$+\overline{\mbox{\bf 10}}$ are found to 
contribute more than this critical amount to the running of the QCD coupling.

These are highly restrictive constraints on the 
construction of a successful GUT scenario containing both VLFs 
and LQs and we see than none of the models discussed above can 
immediately satisfy them {\it unless} the LQ and VLF
superfields can be placed into a single $SU(5)$ representation. In the 
standard $SU(5)$ picture, we can then place $(U,D)^T$, an isosinglet $E^c$ and 
$LQ^c$ into a single {\bf 10} with the corresponding conjugate fields in the 
$\overline{\mbox{\bf 10}}$. This would form a hybrid of model A with the 
`skeleton' model D, which we've denoted by AD above. Of course we pay no 
penalty for also 
including `skeleton' model C here as well, which then yields model ACD.
Instead, when we consider the flipped-$SU(5)\times U(1)$ case, it would appear 
that we can place 
$(N,E)^T$ and $LQ^c$ into a $\overline{\mbox{\bf 5}}$ with the conjugate 
fields in the {\bf 5}; this is exactly model B. It would also seem that 
no penalty is paid as far as unification is concerned for including the 
`skeleton' model C here as well  
{\it except} that this would violate assumption ($iii$) about the chirality of 
LQ couplings to fermions. However, this model is no 
longer truly unified since the hypercharge generator is not fully contained 
within the $SU(5)$ group itself and lies partly in the additional $U(1)$. 
While the $SU(3)_C$ and $SU(2)_L$ couplings will unify, $U(1)_Y$ will not join 
them even when arbitrary additional vector-like singlet fields are added. 
Thus unification no longer occurs in this scenario so that this possibility 
is now excluded. 

The LQ embedding situation becomes more perplexing if the LQ 
and VLFs cannot 
occupy the same GUT multiplet. In this case unification and asymptotic 
freedom constraints become particularly tight and we are forced to consider 
the split multiplet approach mentioned above. This means that we add the 
fields $(2,3)(1/6)\oplus (1,1)(1)\oplus (1,\bar 3)(-2/3)$ and 
their conjugates at low energies but constrain them to be from different 
$SU(5)$ representations. In this case the combination 
$(1,3)(2/3)\oplus (1,\bar 3)(-2/3)$ corresponds to the isosinglet LQ 
and its conjugate so what remains can only be the 
VLF fields. Note that we have again 
arrived back at models AD and ACD. Are these the only solutions? We have 
performed a systematic scan over a very large set of 
VLFs with various 
electroweak quantum numbers under the assumption that they are either color 
singlets or triplets, demanding only that ($i$) QCD remains asymptotically 
free and ($ii$) the model passes the so-called ``B-test''{\cite {mpeskin}} 
which is highly non-trivial to arrange. Essentially the B-test takes advantage 
of the observation that if we know the couplings at the weak scale and we 
demand that unification takes place {\it somewhere} then the values of the 
one-loop beta functions must be related. Note that it is a necessary but not 
sufficient test on our choice of models but is very useful at chopping away 
a large region of parameter space. Using the latest experimental 
data{\cite {moriond}}, we find that
\begin{equation}
B = {b_3-b_2\over {b_2-b_1}} = 0.720\pm 0.030   \,, 
\end{equation}
where the $\pm 0.030$ is an estimate of the corrections due to higher order as 
well as threshold effects and the $b_i$ are the one-loop beta functions of the 
three SM gauge groups. Note that $B_{MSSM}=5/7 \simeq 0.714$ clearly satisfies 
the test. If we require that ($i$) and ($ii$) be satisfied and also require 
that the unification scale not be too low then only the solutions described 
above survive after examining $>7\times 10^{7}$ combinations of matter 
representations. While not completely exhaustive this search indicates the 
solutions above are fairly unique. It is interesting to observe that models 
constructed around model A produce successful grand unification both with and 
without SUSY.

\section{Conclusion and Outlook}

In this talk we have seen how a wealth of data from low and medium energy 
experiments as well as high energy colliders can be combined to point us in a 
fixed direction for LQ model building. I have also discussed a general 
framework for the construction of new $F=0$ scalar 
LQ models which go beyond the original classification by Buchm\"uller, 
R\"uckl and Wyler. This approach is based on the observation that in any 
realistic extension of the SM containing LQs it is expected 
that the LQs themselves will 
not be the only new ingredient. This construction technique is, of course, 
far more general than that required to address the specific issue of the HERA 
excess. While the assumptions of gauge invariance 
and renormalizability are unquestionable requirements of model 
building, it is possible that the other conditions one usually 
imposes are much too strong--unless they are clearly demanded by data. 
This observation implies that for LQs to be experimentally accessible 
their couplings to SM fermions must be essentially 
chiral and separately conserve both Baryon and Lepton numbers. The assumption 
that LQs couple to only a single 
SM generation is surely a convenient way of avoiding numerous low energy 
flavor changing neutral current constraints but is far from natural in the mass 
eigenstate basis. What is required to obtain a new class of LQ models is that 
the LQs themselves must be 
free to couple to more than just the SM fermions and gauge fields. 

Given the fixed gauge structure of the SM the most likely new interactions 
that LQs may possess are with the Higgs field(s) responsible for spontaneous 
symmetry breaking and with new VLFs that are a common 
feature in many extensions of the SM. In the discussion above it has been 
shown how two new forms of the effective interactions of LQs with 
the SM fermions, consistent with Tevatron searches, the HERA excess in both 
the NC and CC channels and low-energy data, can arise through the action of 
VLFs and ordinary symmetry breaking.  The typical VLF mass was found 
to lie in the low TeV region and they could thus be directly produced at 
future colliders with known rates.

We saw that we could construct 
ten new models which fell into two broad classes according to the chirality of 
the resulting LQ couplings to the SM fermions. The VLFs themselves were shown 
to lead to a number of model-dependent effects which are close to the boundary 
of present experimental sensitivity. 
LQs within the framework of models containing VLFs 
were also shown to be consistent 
with Grand Unification in both a supersymmetric {\it and}~non-supersymmetric 
context. The common feature of both schemes is the structure associated with 
model A, \ie, the VLFs are color triplet, weak isodoublets 
in a $(2,3)(1/6)$ 
representation and both $H$ and $H^c$ Higgs fields are required to be 
present as is $LQ^c$ field. In both scenarios the GUT scale is raised 
appreciably from the corresponding model wherein LQs and vector-like 
fermions are absent. In the SUSY case a $(1,1)(1)$ field is also 
required with the optional addition of a SM singlet, corresponding to models 
AD and ACD. In some sense, ACD is the ``anti-$E_6$'' model in that the 
color triplet VLFs are in isodoublets while the color singlet 
fields are all 
isosinglets. Interestingly, in this scenario there is a vector-like 
fermion corresponding to every type of SM fermion. 

Realistic LQ models provide a rich source of new physics beyond the
Standard Model.

\section*{Acknowledgments}
The author would like to thanks J.L. Hewett for collaboration on the work 
on which this talk is based. The author also appreciates comments, 
discussions and 
input from S. Eno(D0), G. Landsberg(D0), J. Conway(CDF) and H. Frisch(CDF) 
regarding the current Tevatron constraints. In addition, he thanks 
Y. Sirois(H1) and D. Krakauer(ZEUS) for discussions of the HERA data.

\newpage
\def\MPL #1 #2 #3 {Mod. Phys. Lett. {\bf#1},\ #2 (#3)}
\def\NPB #1 #2 #3 {Nucl. Phys. {\bf#1},\ #2 (#3)}
\def\PLB #1 #2 #3 {Phys. Lett. {\bf#1},\ #2 (#3)}
\def\PR #1 #2 #3 {Phys. Rep. {\bf#1},\ #2 (#3)}
\def\PRD #1 #2 #3 {Phys. Rev. {\bf#1},\ #2 (#3)}
\def\PRL #1 #2 #3 {Phys. Rev. Lett. {\bf#1},\ #2 (#3)}
\def\RMP #1 #2 #3 {Rev. Mod. Phys. {\bf#1},\ #2 (#3)}
\def\ZPC #1 #2 #3 {Z. Phys. {\bf#1},\ #2 (#3)}
\def\IJMP #1 #2 #3 {Int. J. Mod. Phys. {\bf#1},\ #2 (#3)}

\end{document}